\documentstyle[prl,aps,floats,epsf,epsfig]{revtex}

\begin{document}

\draft

\title{An Exact Solution to a Three Dimensional 
Ising Model and Dimensional Reductions}

\author{Zohar Nussinov}
\address{Institute Lorentz for Theoretical Physics, Leiden University\\
P.O.B. 9506, 2300 RA Leiden, The Netherlands}
\date{\today ; E-mail:zohar@lorentz.leidenuniv.nl}

\twocolumn[

\widetext
\begin{@twocolumnfalse}

\maketitle

\begin{abstract}

A high temperature expansion is employed to
map some complex anisotropic nonhermitian three 
and four dimensional Ising models with algebraic 
long range interactions into a solvable 
two dimensional variant. We also 
address the dimensional reductions for anisotropic two dimensional XY 
and other models. For the latter and
related systems it is possible to have
an effective reduction in the dimension
{\em without the need of compactifying }some dimensions. Some solutions are presented. 
This framework further allows for some very 
simple general observations. It will be 
seen that the absence of a ``phase interference''
effect plays an important role in high dimensional 
problems. A  very forbidding purely algebraic 
recursive series solution
to the three dimensional nearest neighbor
Ising model will be given. In the aftermath, the full-blown three 
dimensional nearest neighbor Ising model is exactly 
mapped onto a  {\em single} spin
1/2 particle with nontrivial 
dynamics. All this allows 
for a formal high dimensional 
Bosonization. 

\end{abstract}

\vspace{0.5cm}

\narrowtext

\end{@twocolumnfalse}
]

\section{Introduction}

In this article, a high temperature expansion
is employed to examine mappings between problems
of various dimensionality.  By these
reductions we will exactly solve an 
anisotropic two dimensional XY and 
some other models.  It will be 
seen that the absence of a ``phase interference''
effect plays an important role in high dimensional 
problems; this observation 
will allow us to examine various 
high dimensional models by averaging
over newly constructed one dimensional systems.
In turn, this averaging process will formally allow
us to effectively bosonize high dimensional
theories.

In section \ref{a},  I am going to start off by discussing 
a momentum space high temperature 
expansion. Next, by analytically continuing
the momenta and making them complex 
(section \ref{exact}) , quaternion (section
\ref{high}), and of a general matrix form we will
be able to exactly solve a variety 
of two, three, and four dimensional
Ising and $O(n)$ models. Many of 
these models are  artificial 
and correspond to non hermitian 
anisotropic long range 
interactions. Some of the resulting
models will be short ranged and of real 
hermitian form. We will exactly solve
some special simple two dimensional 
anisotropic XY models

We will then extend these techniques to the 
quantum case and discuss 
``2+1 dimensional Bethe 
Ansatz'' solutions in 
section(\ref{Bethe})
where we will also
solve a special two dimensional
anisotropic XY model.

After solving these models, in subsection(\ref{comm})
we will apply the high
temperature expansion to
reduce the three dimensional
nearest neighbor Ising ferromagnet
into an illusory {\em formal} solvable 
expression written in a determinant
notation. In this method, the 
high temperature coefficient to each order 
is arrived at by looking a {\em single linear relation}
originating from a matrix of 
(super exponentially) increasing order.  

In subsection(\ref{incomm}),  we will  
map the nearest neighbor $d$ dimensional 
Ising models to an ensemble average 
over a spin chain with long 
ranged ``Coulomb'' like interaction.
In this scenario each spin  is
effectively composed of $d$ 
constituent charges each generating 
its own shifted ``Coulomb'' like field.
This in turn will enable us
to map the $d$ dimensional Ising
model onto a
a single spin 1/2 quantum particle
in 0+1 dimensions.  
We will apply the dimensional 
reduction techniques to quantum
spin problems where an effective
one dimensional fermion problem 
will result. 

In section(\ref{permutational}), 
I will discuss a ``permutational''
symmetry present in the $O(n \rightarrow
\infty)$ models but violated to 
${\cal{O}}(\beta^{4})$ in finite 
$n$ problem. This will again allow 
us to map any large $n$ problem onto 
a one dimensional system but this time 
without the need of averaging. 

One of the main motifs 
of this article is the 
 absence or presence of
``momentum interference''
effects in high and
low dimensional problems 
respectively.

\bigskip

\section{A momentum space high temperature 
expansion}
\label{a}

We will for the most part examine simple classical 
spin models of the type

\begin{eqnarray}
H = \frac{1}{2} \sum_{\vec{x},\vec{y}}
\hat{V}(\vec{x},\vec{y})[\vec{S}(\vec{x}) 
\cdot \vec{S}(\vec{y})]. 
\end{eqnarray} 

Here, the sites $\vec{x}$ and $\vec{y}$
lie on a (generally hypercubic)  
lattice of size $N$. The spins $\{ S(\vec{x}) \}$
are normalized and have $n$ components. The kernel
$\hat{V}(\vec{x},\vec{y})$ is translation invariant.
The $n=1$ case simply corresponds to the 
Ising model.

\bigskip

Later on, when we will discuss quaternion models
we will replace 
the scalar product by a slightly more
complicated product. 

\bigskip

The Hamiltonian in the (non-symmetrical)
Fourier basis is diagonal 
($f(\vec{k}) = \sum _{\vec{x}} 
F(\vec{x}) e^{-i \vec{k} \cdot \vec{x}};$
\bigskip
$ ~ F(\vec{x}) = \frac{1}{N} \sum_{\vec{k}} 
f(\vec{k}) e^{i \vec{k} \cdot \vec{x}}$) and reads 

\begin{eqnarray}
H =  \frac{1}{2N} \sum_{\vec{k}} \hat{v}(\vec{k}) |\vec{S}(\vec{k})|^{2}
\end{eqnarray}

where $\hat{v}(\vec{k})$ and $\vec{S}(\vec{k})$ are the Fourier
transforms of $\hat{V}(\vec{x})$ and $\vec{S}(\vec{x})$
respectively.

For simplicity, we will set the lattice constant to unity-
i.e.  on a hypercubic lattice (of side $L$) with
periodic boundary conditions (which we will
assume throughout) the wave-vector
components $k_{l} = \frac{2 \pi r_{l}}{L}$
where $r_{l}$ is an integer (and the real space 
coordinates  $x_{l}$ are integers).

In the up and coming $V \equiv \beta \hat{V}$. For 
an invertible $\hat{V}(\vec{x},\vec{y})$,
and $0<\beta<\infty$ (via a Hubbard-Stratonovich 
transformation \cite{Hubbard-Statonovich}):
\begin{eqnarray}
  Z=Tr \exp\{-\frac{1}{2}
  \sum_{\vec{x},\vec{y}}V(\vec{x},\vec{y})S(\vec{x})S(\vec{y})\}
\nonumber 
\\ =
  \sqrt{\det(-\frac{2}{\pi V})}\int\prod_{\vec{x}} d\eta(\vec{x})
  \exp[-H\{\eta(\vec{x})\}]
\end{eqnarray}
where
\begin{eqnarray}
  H[\{\eta(\vec{x})\}] = -\frac{1}{2} \sum_{\vec{x},\vec{y}}
  \eta(\vec{x})V^{-1}(\vec{x},\vec{y})\eta(\vec{y})\nonumber
\\ -\sum_{\vec{x}}\ln[\cosh
  \eta(\vec{x})]
\end{eqnarray}
and analogously for the $O(n>1)$ model
\begin{eqnarray}
  H\{\vec{\eta}(\vec{x})\}= -\frac{1}{2} \sum_{\vec{x},\vec{y}}
  V^{-1}(\vec{x},\vec{y})\vec{\eta}(\vec{x})\cdot \vec{\eta}(\vec{y})
\nonumber
\\ 
  -\sum_{\vec{x}}\ln[~\Gamma(n/2)~
  (\frac{2}{|\vec{\eta}(\vec{x})|})^{n/2-1}~
  I_{n/2-1}(|\vec{\eta}(\vec{x})|)~]
\end{eqnarray}
with  $I_{n/2-1}(z)$ a Bessel function.

The $\cosh \eta$ or Bessel terms
may be viewed as terms of constraint
securing the normalization
of $\vec{S}(\vec{x})$ at 
every site $\vec{x}$.

If a magnetic field $h(\vec{x})$ were
applied, the argument of the $cosh$ 
would be replaced $\cosh(\eta+ h)$ with 
a similar occurrence for the Bessel function
appearing for the  
$O(n>1)$ models.

The $q-$state Potts model 
can be viewed as a spin
model in which the $q$ possible
polarizations of the spins
lie at the vertices of a $(q-1)$ 
dimensional tetrahedron- in this manner
the scalar product amongst
any two non identical spins
is $\cos^{-1}(-1/(q-1))$. Employing this
representation, the sum $\sum_{\vec{x}} \ln \cosh \eta(\vec{x})$
may be replaced by $\sum_{\vec{x}} \ln \sum_{i=1}^{q}
\exp[\vec{S}_{i} \cdot \vec{\eta}]$
where the sum is over the $q$ polarizations
of the $(q-1)$ dimensional spin 
$\vec{S}_{i}$.  

The partition function of the 
Ising spins reads
\begin{eqnarray}
Z=  \sqrt{\det(-\frac{2}{\pi V})}\int D \eta e^{- \beta H} \nonumber
\\ =  \sqrt{\det(-\frac{2}{\pi V})}\int \prod_{\vec{x}} d\eta(\vec{x})
\nonumber
\\ \exp \Big[-\frac{1}{2} \sum_{\vec{x},\vec{y}} \eta(\vec{x})
V^{-1}(\vec{x},\vec{y}) \eta(\vec{y}) \Big] \nonumber
\\ \prod_{\vec{x}} \cosh \eta(\vec{x}).
\end{eqnarray}
For a translationally invariant 
interaction $V(\vec{x},\vec{y}) = V(\vec{x}-\vec{y})$, 
this may be written in momentum space
\begin{eqnarray}
Z=  \sqrt{\det(-\frac{2}{\pi V})}\int 
D \eta \exp \Big[-\frac{1}{2N} \nonumber
\\ \sum_{\vec{k}} v^{-1}(\vec{k})
\eta(\vec{k}) \eta(-\vec{k}) \Big] \nonumber
\\ \prod_{\vec{x}} \sum_{m=0}^{\infty} \frac{\eta^{2m}(\vec{x})}{(2m)!},
\end{eqnarray} 
where $v(\vec{k})$ is the Fourier transform
\begin{eqnarray}
v(\vec{k}) = \sum_{\vec{x}} V(\vec{x}-\vec{y}) e^{ i \vec{k} \cdot
\vec{x}} \equiv \beta \hat{v}(\vec{k}).
\end{eqnarray}
where the sum is over all lattice
sites $\vec{x}$. 
Thus the partition function is , trivially, 
\begin{eqnarray} 
Z =  {\cal{N}} \sum_{m_{1},...m_{N}} \langle 
\prod_{i=1}^{N} \frac{1}{(2 m_{i})!}
 \eta^{2m_{i}}(\vec{x}_{i})\rangle_{0} 
\end{eqnarray}
where $\langle ~ ~ \rangle_{0}$ denotes an average 
with respect to the unperturbed Gaussian weight 
\begin{equation}
\exp \Big[-\frac{1}{2N} \sum_{\vec{k}} v^{-1}(\vec{k})
\eta(\vec{k}) \eta(-\vec{k}) \Big],
\end{equation}
and ${\cal{N}}$ is 
a normalization constant. 
Each contraction $\langle \eta(\vec{x}) \eta(\vec{y}) \rangle$
(or $\langle \eta(\vec{k}) \eta(-\vec{k}) \rangle$)
leads to a factor of  $V(\vec{x},\vec{y}) = \beta
\hat{V}(\vec{x},\vec{y})$  
(or to $\beta \hat{v}(\vec{k})/N$ in momentum 
space). Thus the resultant series 
is an expansion in the
inverse temperature $\beta$. In the up and coming we will 
focus attention on the momentum space formulation
of this series. All the momentum space algebra
presented above is only a slight modification 
to the well known high temperature expansions usually
generated by the Hubbard Stratonovich transformation 
directly applied to the real space representation
of the fields $\eta(\vec{x})$ \cite{Wortis,Polyakov}. 
It is due to the naivete of the author that 
such a redundant momentum space formulation
was rederived in the first place. However, 
as we will see, in momentum space some properties 
of the series become much more transparent. To 
make $V$ invertible and the expansion convergent, 
we will shift $\hat{v}(\vec{k})$ by a constant 
\begin{equation}
\hat{v}(\vec{k}) \rightarrow \hat{v}(\vec{k}) + A
\end{equation}
such that $\hat{v}(\vec{k})$ is strictly positive. 
In real space such a constant shift amounts
to a trivial shift in the on site interaction 
(or chemical potential)
\begin{eqnarray}
\hat{V}(\vec{x},\vec{y}) \rightarrow \hat{V}(\vec{x},\vec{y}) +A.
\delta_{\vec{x},\vec{y}}
\end{eqnarray}
For asymptotically large $|\eta|$ the nontrivial 
\begin{eqnarray}
H^{dual}_{1} \equiv \sum_{\vec{x}} \ln |\cosh \eta(\vec{x})|
\end{eqnarray}
is linear in $|\eta|$. The Gaussian generating
\begin{eqnarray}
H^{dual}_{0} = \frac{1}{2} \sum_{\vec{x},\vec{y}}
  \eta(\vec{x})V^{-1}(\vec{x},\vec{y})\eta(\vec{y})
\end{eqnarray}
is a positive quadratic
form and dominates at large 
$\eta$.  
So far all of this
has been very general.
Now let us consider the
$d$ dimensional nearest 
neighbor Ising model. 
To avoid carrying minus signs around, we will consider 
 an Ising antiferromagnet whose 
partition function is, classically, identically 
equal to that of a ferromagnet. The exchange 
constant $J$ will be set to unity. The corresponding momentum space
kernel 
\begin{equation}
\hat{v}(\vec{k}) =
A + \sum_{l=1}^{d} [e^{i k_{l}} + e^{-i k_{l}}].
\label{d-dim}
\end{equation}

As the reader might guess, the uniform shift $A$ 
only leads to a trivial change in $Z$
(i.e. to a multiplication by the constant 
$\exp[- \beta A N]$).
In the forthcoming we will set $A=0$.

Owing to the 
relation
\begin{equation}
\int_{-\pi}^{\pi} dk ~e^{ikN} = 2 \pi \delta_{N,0}
\label{int}
\end{equation}
all loop integrals are
trivial.
A simple integral
is of the type 

\begin{eqnarray} 
\int_{\sum_{i=1}^{M} \vec{k}_{i}=0}
\prod_{i=1}^{M} \frac{d^{d}k_{i}}{(2 \pi)^{d}} \prod_{i} 
v^{n}(\vec{k}_{i}) \nonumber
\\ = \sum_{n=\sum_{l=1}^{d} (\gamma_{l}+\delta_{l})} [ \frac{n!}{\prod_{l=1}^{d} \gamma_{l}! \delta _{l}!}]^{M}
\end{eqnarray}
(when $n=1$ this integral is $2d$ etc).
A related integral reads 
\begin{eqnarray}
\int \frac{d^{d}k}{(2 \pi)^{d}}
\prod_{\sum_{i=1}^{M} \vec{k}_{i}=0}~~
[v(\vec{k}_{i})]^{n_{i}} \nonumber
\\ = \sum_{\gamma_{l}^{i}-\delta_{l}^{i}
= m_{i}^{l},~\sum_{i=1}^{M} m_{i}^{l} =0,~ 
\sum_{l=1}^{d} \gamma_{l}^{i}+\delta_{l}^{i} = n_{i}}
\prod_{i=1}^{M} [\frac{n_{i}!}{\prod_{l=1}^{d} \gamma_{l}^{i}! \delta _{l}^{i}!}]
\end{eqnarray}
where in the last sum $m_{j<n}^{l} \neq m_{i <M}^{l} = -m_{i=M}^{l}$
etc. 
Note that the high temperature expansion 
becomes simpler if instead of symmetrizing
the interaction (each bond being counted twice-
once by each of the two interacting spins) 
one considers 
\begin{equation}
\hat{v}(\vec{k}) = 2 \sum_{l=1}^{d} \exp[i k_{l}].  
\end{equation}
Here each spin interacts with only $d$ (and not $2d$) of
its neighbors separated from it 
by one positive distance along
the d axes $\hat{e}_{l=1}^{d}$. 
The factor of 2 originates as each 
bond is now counted only once
and therefore the corresponding bond 
strength is doubled.
The partition function Z is 
trivially unchanged. 

In the most general diagram
the propagator momenta $\{ \vec{q}_{a} \}$
($1 \le a \le$ number of propogators)
are linear combinations of the 
independent loop momenta $\{ \vec{k}_{b} \}_{b=1}^{\mbox{loops}}$
\begin{eqnarray}
\vec{q}_{a} = M_{ab} \vec{k}_{b}.
\end{eqnarray}

Symmetry factors aside,
the value of a given 
diagram reads
\begin{eqnarray}
\int \prod_{b} \frac{d^{d}k_{b}}{(2 \pi)^{d}} \prod_{a} v(\Sigma M_{ab}
k_{b})  \nonumber
\\ = \int \prod_{b} \frac{d^{d}k_{b}}{(2 \pi)^{d}} \prod_{a} (2 \beta)
\Sigma_{l=1}^{d}  \exp[i \Sigma_{b} M_{ab} k_{b}^{l}] 
\nonumber
\\ =  (2 \beta)^{\mbox{propagators}} \nonumber
\\ \times \prod_{b=1}^{\mbox{loops}} \Big( \int \frac{d^{d}k_{b}}{(2 \pi)^{d}}
\Sigma_{l=1}^{d}  \exp[i \sum_{a,b} M_{ab} k_{b}^{l}] \Big)
\label{loop}
\end{eqnarray}
- i.e. upon expansion of the outer sum, 
just a product of individual
loop integrals of the type encountered
in Eqn.(\ref{int}) for each $k^{l}_{b}$.

The latter integral in $d=1$ reads
\begin{eqnarray} 
I = \prod_{b=1}^{\mbox{loops}} \delta(\Sigma_{a=1}^{\mbox{propagators}} 
M_{ab},0)
\end{eqnarray}
The situation in $d >1$ becomes far richer. 
One may partition the integral into all
possible subproducts of momenta corresponding
to different spatial components $l$ satisfying
the delta function constraints.
In a given diagram not all propagators need
to correspond to the same spatial 
component $k^{l}$. The combinatorics of
``in how many ways loops may be chosen
to satisfy the latter delta
function constraints?'' boils 
down to the standard counting 
of closed loops in real 
space. 

Some simple technical details of 
lattice perturbation theory
vis a vis the standard continuum
theories are presented in 
\cite{N}.

\section{ An exact solution
to a three dimensional Ising
model}
\label{exact}

{\bf Theorem:} 

If the spin spin interaction on a cubic lattice 
has a non hermitian kernel of
the form 
\begin{eqnarray}
\hat{V}(x_{1},x_{2},x_{3}) = - \delta_{x_{2},0}
\delta_{x_{3},0}  (\delta_{x_{1},1}+ \delta_{x_{1},-1}) 
\nonumber
\\ - \frac{1}{\pi} (-1)^{x_{3}} \frac{\sinh \pi \lambda}{\lambda^{2}+ x_{3}^{2}}
 \delta_{x_{1},0} \Big[\lambda  (\delta_{x_{2},1}+ \delta_{x_{2},-1}) \nonumber
\\ + i x_{3} ( \delta_{x_{2},1}- \delta_{x_{2},-1})] 
 \Big]
\end{eqnarray}
then, for all real $\lambda$,
the Helmholtz free energy per spin
$a(h=0,T)$ is
\begin{eqnarray}
\beta f = - \ln( 2 \cosh 2 \beta) -
\nonumber
\\ 
\frac{1}{2 \pi} \int_{0}^{\pi} d \phi \ln \frac{1}{2}
(1+ \sqrt{1 - \kappa^{2} \sin^{2} \phi})
\end{eqnarray}
where
\begin{eqnarray}
\kappa \equiv \frac{2 \sinh 2 \beta}{\cosh^{2} 2 \beta}
\end{eqnarray}
and the internal energy per spin
\begin{eqnarray}
u=  - \coth 2 \beta  \Big[ 1+ \frac{2}{\pi}
\kappa^{\prime} K_{1}(\kappa) \Big]
\end{eqnarray}
where $K_{1}(\kappa)$ is a complete 
integral of the first kind
\begin{eqnarray}
K_{1}(\kappa) \equiv \int_{0}^{\pi/2} \frac{d \phi}{\sqrt{1
-\kappa^{2} \sin^{2} \phi}}
\end{eqnarray}
and 
\begin{equation}
\kappa^{\prime} \equiv 2 \tanh^{2} (2 \beta)  -1.
\end{equation}

The proof is quite simple. The
expressions presented are
the corresponding intensive quantities for 
the two dimensional nearest neighbor Ising model\cite{Onsager}.

Let us start by writing down the 
diagrammatic expansion for
the two dimensional Ising
model and set
\begin{eqnarray}
\hat{v}(q_{1},q_{2}) \rightarrow \hat{v}(q_{1},q_{2}+ i \lambda q_{3})
\end{eqnarray}
in all integrals. This effects
\begin{eqnarray}
\int_{-\pi}^{\pi} dk_{2} e^{i k_{2}N^{loop}} \rightarrow \int_{-\pi}^{\pi}
dk_{2}
e^{ik_{2}N^{loop}} \int_{-\pi}^{\pi} dk_{3} e^{-\lambda k_{3} N^{loop}}
\end{eqnarray}
for all individual loop integrals.
The integrals are nonvanishing only 
if $\{N_{b}=0 \}$ for all loops $b$. 
When $N^{loop}=0$ for a given loop
the second ($k_{3}$ integration)
leads to a trivial multiplicative
constant (by one). In the original
summation 
\begin{eqnarray}
\sum_{\vec{k}_{b}} \rightarrow \frac{L^{d}}{(2 \pi)^{d}} \int_{-\pi}^{\pi} ... 
\int_{-\pi}^{\pi} d^{d}k_{b}
\end{eqnarray}
and thus  an additional 
factor of $L$ is introduced for each independent connected
diagram. Thus the free energy per unit volume 
for a system with the momentum
space kernel $\hat{v}(k_{1},k_{2}+i \lambda k_{3})$
is the same as for the two dimensional
nearest neighbor ferromagnet with
the kernel $\hat{v}(k_{1},k_{2})$.
Fourier transforming 
 $\hat{v}(k_{1},k_{2}+i \lambda k_{3})$
and symmetrizing $[V(\vec{x}) + V(-\vec{x})]/2 \rightarrow V(\vec{x})$
one finds the complex real space
kernel $\hat{V}(x_{1},x_{2},x_{3})$
presented above \cite{Gibbs}
Q.E.D.

\bigskip

{\bf Corollary}

For the more general 
\begin{eqnarray}
\hat{V}(x_{1},x_{2},x_{3}) = -\delta_{x_{2},0} \delta_{x_{3},0}
(\delta_{x_{1},1} + \delta_{x_{1},-1}) \nonumber
\\ 
- {\cal{J}} \delta_{x_{1},0}\Big[\frac{(-1)^{x_{3}}}{\lambda^{2}+ x_{3}^{2}}\Big] 
\{ (\delta_{x_{2},1}+ \delta_{x_{2},-1})  \nonumber
\\ + i
(\delta_{x_{2},-1}-\delta_{x_{2},1}) \frac{x_{3}}{\lambda} \}
\label{Neel}
\end{eqnarray}
we may define 
\begin{eqnarray}
r \equiv \frac{\pi {\cal{J}}}{2 \lambda ~\sinh \pi \lambda}
\end{eqnarray}
to write the free energy
as the free energy of an 
anisotropic two dimensional 
nearest neighbor ferromagnet
with the kernel
\begin{eqnarray}
\hat{V}(x_{1},x_{2})= -\delta_{x_{2},0} (\delta_{x_{1},1} +
\delta_{x_{1},-1} \nonumber
\\ - r \delta_{x_{1},0}(\delta_{x_{2},1} + \delta_{x_{2},-1})
\end{eqnarray}
i.e. a nearest neighbor Ising ferromagnet having the 
ratio of the exchange constants along the $x_{1}$ and $x_{2}$
axes equal to $r= J_{2}/J_{1}$ (throughout we set $J_{1}$ to unity).
Introducing
\begin{eqnarray}
u \equiv \frac{1}{\sinh 2 \beta \sinh 2 \beta r}
\end{eqnarray}
and defining 
\begin{eqnarray}
F(\theta) \equiv \ln \{2 [\cosh 2 \beta \cosh 2 \beta r \nonumber
\\ + u^{-1} (1- 2 u
\cos \theta + u^{2})^{1/2}]\},
\end{eqnarray}
we may write the free energy density
as
\begin{equation}
\beta f = - \frac{1}{2 \pi} \int_{0}^{\pi} F(\theta).
\end{equation}

\bigskip

\bigskip

Note that the same result 
applies for the ``ferromagnetic''

\begin{eqnarray}
\hat{V}(x_{1},x_{2},x_{3}) = -\delta_{x_{2},0} \delta_{x_{3},0}
(\delta_{x_{1},1} + \delta_{x_{1},-1}) \nonumber
\\ 
- {\cal{J}} \delta_{x_{1},0}\Big[\frac{1}{\lambda^{2}+ x_{3}^{2}}\Big] 
\{ (\delta_{x_{2},1}+ \delta_{x_{2},-1})  \nonumber
\\ + i
(\delta_{x_{2},-1}-\delta_{x_{2},1}) \frac{x_{3}}{\lambda} \}.
\label{ferro}
\end{eqnarray}
The proof is simple- the partition function
is identically the same if evaluated with the
kernel in Eqn.(\ref{ferro}) instead
of that  in Eqn.(\ref{Neel})
if the spins are flipped
\begin{eqnarray}
S(\vec{x}) \rightarrow (-1)^{x_{3}} S(\vec{x}).
\end{eqnarray}

\bigskip

\bigskip

What about correlation functions?
The equivalence of the partition
functions $Z[\beta,h(x_{1},x_{2})]$
implies that in the 
planar directions
the correlations
are the same as
for two dimensional ferromagnet.

Changing the single momentum
coordinate
\begin{eqnarray}
v(k_{1},k_{2}) \rightarrow v(k_{1},k_{2}+ i \lambda k_{3}) \nonumber
\end{eqnarray}
effects 
\begin{eqnarray}
G(k_{1},k_{2}) \rightarrow G(k_{1},k_{2}+ i \lambda k_{3})
\label{cor}
\end{eqnarray}
in the momentum space correlation function.
The algebra is identically the same.

The connected spatial correlations (for $T \neq T_{c}$) 
are exponentially  damped 
along the $x_{2}$ axis 
as they are in the usual two dimensional 
nearest neighbor ferromagnet (see \cite{Wu,Mc} 
for the two dimensional correlation
function).

Of course, all of this can also be extended
to one-dimensional like spin systems.

The two dimensional spin-spin kernel
\begin{eqnarray}
\hat{V}(x_{1},x_{2}) =   - \frac{1}{\pi} (-1)^{x_{2}} \frac{\sinh \pi \lambda}{\lambda^{2}+ x_{2}^{2}}
 \delta_{x_{1},0} \Big[\lambda  (\delta_{x_{2},1}+ \delta_{x_{2},-1}) \nonumber
\\ + i x_{2} ( \delta_{x_{2},1}- \delta_{x_{2},-1})] 
 \Big]
\label{special}
\end{eqnarray}
leads to one dimensional behavior
with oscillatory oscillations 
along the $x_{2}$.
More precisely, the  momentum space
correlator for any $O(n)$ system in
one dimension reads
\begin{eqnarray}
G(k) =  \frac{e^{r}}{e^{r}-e^{ik}} + \frac{1}{e^{ik+r}-1}.
\end{eqnarray}
For the Ising ($n=1$) system $r \equiv - [\ln \tanh \beta]$.
One may set
$k = k_{1} + i \lambda k_{2}$ and 
compute the inverse 
Fourier transform.
The correlations along 
$x_{2}$ are oscillatory
with temperature dependent 
wavevectors.  In other
words, the effective correlation length
along $x_{2}$ is infinite.

If the two dimensional spin-spin interactions
in Eqn.(\ref{special}) are augmented by an additional 
on-site magnetic field $h$ then the two dimensional 
partition function reads 
\begin{equation}
Z = \lambda_{+}^{N}+ \lambda_{-}^{N}
\end{equation}
where, as usual, 
\begin{eqnarray}
\lambda_{\pm} = e^{\beta} \cosh \beta h \pm \sqrt{e^{2 \beta}
\sinh^{2}\beta h + e^{-2 \beta}}.
\end{eqnarray}

That ``complexifying'' the coordinates 
should not change the physics is intuitively
obvious: if one shifts $k_{1} \rightarrow k_{1} + \lambda k_{2} \equiv
k^{\prime}_{1}$
with real $\lambda$ in the one dimensional 
$\hat{v}(k_{1})$ then the resulting 
kernel is $\hat{v}(k^{\prime}_{1})$ plainly describes 
a stack on chains parallel to $(1,\lambda)$; the 
spins interact along the chain direction yet
the chains do not interact amongst themselves. 
The free energy density should be identically
equal to that of a one dimensional 
system- no possible dependence on 
$\lambda$ can occur. All that we have 
done in the above is allow
$\lambda$ to become complex.

In 
\begin{eqnarray}
\hat{v}(k_{1},k_{2}) \rightarrow \hat{v}(k_{1},k_{2}+ \lambda k_{3})
\end{eqnarray}
the correlation functions along
a direction orthogonal to the 
direction $(1,\lambda)$ in the $(x_{2},x_{3})$
plane vanish. If $\lambda$ is complex 
then there are no correlations along an
orthogonal direction in the 
``complex'' space.

Formally, all this stems from the trivial fact
that the measure $dk_{1} dk_{2} = const (dk dk^{*})$,
whereas $\hat{v}(k)$ depends only on the 
single complex coordinate $k$.

Other trivial generalizations
are possible. For example, we 
may make both $k_{1}$ and $k_{2}$
complex in $\hat{v}(k_{1},k_{2})$
to generate a four dimensional spin kernel
\begin{eqnarray}
\hat{V}(x_{1},x_{2}.x_{3},x_{4}) \nonumber
\\=
 - \frac{1}{\pi} \Big( (-1)^{x_{2}} \frac{\sinh \pi \lambda_{1}}{\lambda_{1}^{2}+ x_{2}^{2}}
 \delta_{x_{1},0} \Big[\lambda_{1}  (\delta_{x_{2},1}+ \delta_{x_{2},-1}) \nonumber
\\ + i x_{3} ( \delta_{x_{2},1}- \delta_{x_{2},-1})] 
 \Big] \nonumber
\\ +  (-1)^{x_{4}} \frac{\sinh \pi \lambda_{2}}{\lambda_{2}^{2}+ x_{4}^{2}}
 \delta_{x_{3},0} \Big[\lambda_{2}  (\delta_{x_{4},1}+
\delta_{x_{4},-1}) 
\nonumber \\ + i x_{4} ( \delta_{x_{3},1}- \delta_{x_{3},-1})] 
 \Big]  \Big)
\end{eqnarray}

which leads to the canonical (nearest neighbor) 
two dimensional behavior for arbitrary 
$\lambda_{1}$ and $\lambda_{2}$.

All of this
need not be restricted
to cubic lattice models. We may also
analytically continue $\hat{v}(\vec{k})$
of the triangular lattice etc.

For the triangular antiferromagnet/ferromagnet
\begin{eqnarray}
v(\vec{k}) = \pm 2[\cos k_{1} + \cos (\frac{{k}_{1}}{2} +
\frac{\sqrt{3}}{2} k_{2}) \nonumber
\\ + \cos(-\frac{k_{1}}{2}+\frac{\sqrt{3}}{2}k_{2})].
\end{eqnarray}
Making $k_{2}$ complex leads to interactions
on a layered triangular lattice etc.

\section{Higher  $O(n>1)$ models}
\label{high}

The multi-component $O(n>1)$ spin
models display a far richer 
variety of possible higher
dimensional 
extensions.

$\lambda$ need not be only a
complex number, it may also be 
quaternion or correspond 
to more general set of matrices.

Let us examine the extension 
to ``quaternion'' $\vec{k}$
(or with the matrices
$i\sigma_{1},i\sigma_{2}$,and $i\sigma_{3}$
(with $\{ \sigma_{\alpha} \}$ the Pauli matrices) 
taking on the role of $i,j$ and $k$).
If we were dealing with an $O(2)$ (or XY) model,
the same high temperature expansion 
could be reproduced:

We may envisage a trivial extension to the usual 
scalar product Hamiltonian
\begin{equation}
H= \frac{1}{2}\sum_{\vec{x},\vec{y}} \hat{V}(\vec{x}-\vec{y})
\vec{S}(\vec{x}) \cdot \vec{S}(\vec{y})  
\end{equation}
to one in which the kernel is no
longer diagonal in the internal 
spin indices $\alpha,\beta  = 1,2$.

\begin{equation}
H= \frac{1}{2}\sum_{\vec{x},\vec{y}} \hat{V}_{\alpha,\beta}(\vec{x}-\vec{y})
S_{\alpha}(\vec{x})  S_{\beta}(\vec{y}).  
\end{equation}

The high temperature expansion may be reproduced as before with the kernel 
$V_{\alpha,\beta}(\vec{x}-\vec{y})$ being generated by each contraction of 
$\eta_{\alpha}(\vec{x})$ and $\eta_{\beta}(\vec{y})$,
or in momentum space $v_{\alpha,\beta}(\vec{k}) 
\delta_{\vec{k}+\vec{k}^{\prime},0}$ 
is generated  by the contraction of 
$\eta_{\alpha}(\vec{k})$ and 
$\eta_{\beta}(\vec{k}^{\prime})$.

Let us start off with a nearest neighbor one dimensional 
XY chain and consider the transformation
\begin{eqnarray}
\exp[i k_{1}] \rightarrow \exp[i (k_{1} + i 
\lambda_{1} k_{2} \sigma_{1} + i \lambda_{2} k_{3} \sigma_{2}
+ i \lambda_{3} k_{4} \sigma_{3})] \nonumber
\\ \equiv \exp[i k_{1} - \vec{w} 
\cdot \vec{\sigma}].
\end{eqnarray}
In the argument of the exponential the identity matrix 
$1$ commutes with $\vec{w} \cdot
\vec{\sigma}$ and the 
exponential may be trivially
expanded as 
\begin{eqnarray}
\exp[i k_{1} - \vec{w} \cdot
\vec{\sigma}] = \exp[i k_{1}] \exp[-  \vec{w} \cdot
\vec{\sigma}].
\end{eqnarray}
Once again in higher dimensions 
(this time $d=4$) the $k_{1}$ integration will
reproduce the familiar 
\begin{equation}
\int_{-\pi}^{\pi} dk_{1} \exp[i N k_{1}] = 2 \pi \delta_{N,0}
\end{equation}
for each independent loop momentum $\vec{k}$
and the integration over the remaining 
$k_{2},k_{3},$ and $k_{4}$ components
will simply yield multiplicative 
constants. 

Thus the kernel  
\begin{equation}
v(\vec{k}) = \exp[i k_{1} - \vec{w} \cdot
\vec{\sigma}] + h.c. 
\end{equation}
will generate a four dimensional
XY model which has the partition 
function of a nearest neighbor one dimensional 
XY chain. The corresponding 
real space kernel $\hat{V}(x_{1},x_{2},x_{3},x_{4})$
\begin{eqnarray}
\hat{V}_{\alpha \beta}
(x_{1},x_{2},x_{3},x_{4}) = \int_{-\pi}^{\pi} ... \int_{-\pi}^{\pi} 
\frac{d^{4}k}{(2 \pi)^{4}} \exp[i \vec{k} \cdot \vec{x}]  
\Big(\exp[i (k_{1}  \nonumber
\\ + i \lambda_{1} k_{2} \sigma_{1} + 
i \lambda_{2} k_{3} \sigma_{2} 
 + i \lambda_{3} k_{4} \sigma_{3})]_{\alpha \beta} + h.c. \Big).
\nonumber
\end{eqnarray}
In general, this kernel is no longer 
diagonal in the internal 
spin coordinates (unless $\vec{w}$ happens
to be oriented along $\sigma_{3}$).
\begin{eqnarray}
|\vec{w}| = \sum_{i=2}^{4} \lambda_{i-1}^{2} k_{i}^{2} \nonumber
\\ \hat{w} = (\frac{\lambda_{1} k_{2}}{|\vec{w}|},  \frac{\lambda_{2}
k_{3}}{|\vec{w}|}, \frac{\lambda_{3} k_{4}}{|\vec{w}|})
\end{eqnarray}
and 
\begin{eqnarray}
\exp[-\vec{\sigma} \cdot \vec{w}] = \cosh (|\vec{w}|) - \sinh
(|\vec{w}|) (\vec{\sigma} \cdot \hat{w})
\end{eqnarray}
For general $\{ \lambda_{i} \}_{i=1}^{3}$ 
the inverse Fourier transform is nontrivial.
The scalar product is replaced
by a more complicated product
amongst the components. 
The real space interaction
kernel is once again
algebraically long ranged
along the $x_{2},x_{2}$ and $x_{4}$
axes.

In this manner
we may also generate
an infinite number
of solvable models
with a {\bf real hermitian}
kernel. We may state 
another

{\bf{Theorem}}

The hermitian kernel
\begin{eqnarray}
v(\vec{k}) = -2 \cos(k_{1} +  k_{2} \sigma_{3})
\label{O_2}
\end{eqnarray}
leads to the partition function of 
a one dimensional $O(2)$ chain.

\bigskip

The proof amounts to a reproduction of the
calculations given above for the four 
dimensional $O(2)$ system.

\bigskip

In real space the latter kernel (Eqn.(\ref{O_2}))
leads to the $O(2)$ 
Hamiltonian 
\begin{eqnarray}
H = - \sum_{\langle i j \rangle~ along~(1,1)} S_{i}^{(1)} S_{j}^{(1)} 
 - \sum_{\langle i j \rangle~ along~(1,-1)} S_{i}^{(2)} S_{j}^{(2)} 
\label{dec}
\end{eqnarray}
where the $n=1$ (or x component) 
of the spins interact in the first 
term and only the y-components
of the XY spins appear in the 
second term; the indices $i$ and
$j$ are the two dimensional 
square lattice coordinates.
The first term corresponds to
interactions along the $(1,1)$
direction in the plane
and the second term corresponds
to interactions along the 
$(1,-1)$ diagonal. 
The two spin components
satisfy $[S_{i}^{(1)}]^{2}+ [S_{i}^{(2)}]^{2}=1$-
i.e. are normalized at every site $i$.
By our mapping, this model 
trivially has the partition function
of a one dimensional nearest neighbor $O(2)$ spin 
chain.

Similarly, the partition
function for the three dimensional
kernel $\hat{v}(\vec{k}) = -2 \cos (k_{1}+k_{2}+k_{3} \sigma_{3})$
is that of a two dimensional nearest neighbor XY model
exhibiting Kosterlitz-Thouless like behavior.
Here each of two spin components interacts within a different
subplane. The coupling between the two spin 
components due to the normalization
constraint apparently plays no role
in leaving the system two dimensional.

The reader can easily 
see how a $d^{\prime}$
dimensional system can
be made have an effective
dimensionality $1 \le d \le d^{\prime}$
without the physical need of
actual compactification.

Polarization dependent hopping (or ``interactions'') 
akin to those in Eqn.(\ref{dec}) may be of relevance in discussing 
directed orbital problems 
in two dimensions.

\section{Special Two Dimensional Bethe Ansatz solutions}
\label{Bethe}

Let us also write the general
Hubbard Stratonovich transformation
for a Quantum system \cite{N-O}.

Consider a system of fermions  with 
the Hamiltonian
\begin{eqnarray}
H = H_{0} + V \nonumber
\\
H_{0} = \sum_{\alpha \beta} T_{\alpha \beta}  a^{\dagger}_{\alpha}
a_{\beta} \nonumber \\
V = \frac{1}{2} \sum_{\alpha \beta \gamma \delta}
\langle \alpha \beta | v | \gamma \delta \rangle 
a^{\dagger}_{\alpha} a^{\dagger}_{\beta} a_{\delta} a_{\gamma} 
\end{eqnarray}

Following Negele and Orland \cite{N-O} it is 
convenient to introduce
\begin{eqnarray}
\hat{\rho}_{\alpha \gamma} = a^{\dagger}_{\alpha} a_{\gamma} \nonumber
\\ K_{\alpha \delta} = T_{\alpha \delta} - \frac{1}{2} \sum_{\beta}
v_{\alpha \beta \beta \delta}
\end{eqnarray}
where, for our applications, $\vec{\alpha}, \vec{\beta},\vec{\gamma},$
and $\vec{\delta}$ are two dimensional position
vectors.

For this time independent Hamiltonian 
\begin{eqnarray}
\exp[- i H(t_{f}-t_{i})] \nonumber
\\ = \int \prod_{\alpha \beta} 
D \eta_{\alpha \beta}(t) \exp[\frac{i}{2} \int_{t_{i}}^{t_{f}} dt
\eta_{\alpha \beta} (t) v_{\alpha \beta \gamma \delta}
\eta_{\gamma \delta} (t) ] 
\nonumber
\\ T \Big[ \exp[-i \int_{t_{i}}^{t_{f}} dt (
K_{\alpha \beta} + \eta_{\gamma \delta} (t) v_{\gamma \alpha \delta
\beta})
\hat{\rho}_{\alpha \beta}] \Big]
\end{eqnarray}

Again a momentum space expansion 
is possible in $\eta_{\alpha \beta} (t)$. 
If an exact Bethe Ansatz solution is known 
for nearest neighbor
hopping then in 
some cases it may
be extended
to two dimensions where the hopping
matrix element $K(x_{1},x_{2})$
for separation $\vec{x}=(x_{1},x_{2})= \vec{\alpha}-\vec{\beta}$
attains exactly the same form as the two-dimensional
kernel $\hat{V}(x_{1},x_{2})$ just given previously.
More generally,  the Trotter formula may be employed.

Loosely speaking, if in a fictitious electronic system
the hopping matrix element would be the non hermitian 
\begin{eqnarray}
K(x_{1},x_{2}) = 
 - \frac{1}{\pi} \Big( (-1)^{x_{2}} \frac{\sinh \pi \lambda_{1}}{\lambda_{1}^{2}+ x_{2}^{2}}
 \delta_{x_{1},0} \nonumber
\\ 
\Big[\lambda_{1}  (\delta_{x_{2},1}+ \delta_{x_{2},-1}) \nonumber
\\ + i x_{3} ( \delta_{x_{2},1}- \delta_{x_{2},-1})
 \Big] \Big)
\end{eqnarray}
then the system would essentially
a one dimensional Luttinger liquid
along $x_{1}$. If the kernel would instead decay 
algebraically along the imaginary 
time axis the problem would become 
that of dissipating system.

A hermitian hopping matrix 
element could also do the trick if we 
were to consider polarization
dependent hopping (or ``interactions'') 
analogous to those in Eqn.(\ref{dec})
and slightly more complicated
variants. As noted earlier, 
such models may be of relevance 
in discussing directed
orbital problems in two 
dimensions.

For mapping the two dimensional $S=1/2$ Quantum problem
of Eqn.({\ref{dec}) onto a nearest neighbor
XY chain we find that the free energy 
density is given by 
\begin{eqnarray}
\beta f = - \int_{0}^{1} dx ~ \ln[1+ \exp(4 \beta \cos (2
\pi x))].
\end{eqnarray}

\section{Phase interference and a mapping onto
a single spin problem}

We will now map high dimensional ($d>1$) problems onto 
translationally invariant problems 
in one dimension. 

For the nearest neighbor 
ferromagnet, the basic idea 
is to make a comparison between the 
the kernels
\begin{eqnarray}
\hat{v}_{d} = -2 \sum_{l=1}^{d} \cos k_{l} \nonumber
\\
\hat{v}_{1} = -2 \sum_{l=1}^{d} \cos c_{l} k_{1}
\label{band}
\end{eqnarray}
in $d$ and in one spatial dimensions respectively. 
Note that the one dimensional
problem has, in general, $d$ 
different harmonics of the single
momentum coordinate $k_{1}$. If the
coefficients $c_{l}$ are integers
equal to $R_{l}$ then the one dimensional 
problem amounts to a ferromagnetic 
chain in which each spin 
interacts with $d$ other
spins at distances $\{R_{l}\}_{l=1}^{d}$
away. When evaluating the loop integrals,
we will find that the partition function/
free energy deviate from their
$d$ dimensional values due to ``interference''
between the various harmonics in 
$\hat{v}_{1}(k_{1})$. If all harmonics
acted independently then the $d$ 
dimensional result would be reproduced.
The advantages of the form $\hat{v}_{1}$ are obvious.
Perhaps the most promising  alley is that of ``incommensurate dimensional
reduction''. By this method, we may examine
the d-dimensional problem described by $\hat{v}_{d}$ by 
a bosonization of the one-dimensional
problem of $\hat{v}_{1}$. A 
solution to the one dimensional problem
posed by $\hat{v}_{1}$ followed by
an average over the coefficients $\{c_{l}\}$
will immediately lead to the 
corresponding $d-$ dimensional 
quantities.

\subsection{Commensurate dimensional reduction} 
\label{comm}

Consider a one dimensional ferromagnetic lattice model 
with a nearest neighbor (or $Range=1$) interaction 
and a $Range=n$ interaction: 
$\hat{v}(k_{1}) = -2  (\cos k_{1}+ \cos nk_{1})$.
The loop integrals 
containing terms of the $\cos k_{1}$ 
origin and terms stemming from 
$\cos nk_{1}$ are independent
to low orders.
Mixed terms can survive only
to orders $(1/T)^{n+1}$
and higher. The delta function
constraints $\sum q_{1} =0$
would generate terms identical
to those of the two 
dimensional nearest neighbor 
model with $\hat{v}(k_{1},k_{2})= -2 (\cos k_{1} + \cos k_{2})$
with the two delta function constraints for
the two separate components of the momentum, 
$\sum q_{1}=0$ and $\sum q_{2}=0$, at 
every vertex. Thus, in a sense, this
system is two dimensional
up to $T^{-n}$. Similarly,
if a one dimensional 
chain has interactions
of length one lattice
constant, $R=n$ and $R=m$
i.e.
\begin{equation}
\hat{v}(k_{1}) = - 2  (\cos k_{1} + \cos n k_{1} + \cos m k_{1})
\label{3-d}
\end{equation}
then this system, as fleshed out
by a $1/T$ expansion for 
the partition function
or for the free energy, is three 
dimensional up to $order ~ = ~ n$
if $m=n^2$.  (If
$m=n+1$ then the
triangular ferromagnet 
is generated). Similar, higher
dimensional extensions
similarly follow from 
the lack of commensurability
of the cosine arguments
(for four dimensions  
\begin{equation}
\hat{v}(k_{1}) = - 2  (\cos k_{1} + \cos n k_{1} + \cos m k_{1} + \cos s k_{1})
\end{equation}
with $s= n^2$ etc.)

The corresponding transfer
matrices are trivial to 
write down. For the three dimensional case:
\begin{eqnarray}
\langle s_{1} ... s_{n^2} | T| s^{\prime}_{1} ... s^{\prime}_{n^2}
\rangle = \exp[\beta\{ \frac{1}{2}(s_{1}s_{2}+ ...+s_{n^2-1}s_{n^2}) \nonumber
\\ +\frac{1}{2} (s^{\prime}_{1}s^{\prime}_{2}+ 
...+s^{\prime}_{n^2-1}s^{\prime}_{n^2}) \nonumber
\\ +
\frac{1}{2} (s_{1}s_{n+1}+s_{2}s_{n+2}+ ...+s_{n^2-n}s_{n^2})
\nonumber
\\ + \frac{1}{2} (s^{\prime}_{1}s^{\prime}_{n+1}+s^{\prime}_{2}
s^{\prime}_{n+2}+ ...+s^{\prime}_{n^2-n}s^{\prime}_{n^2}) \nonumber
\\ + s_{n^{2}}s^{\prime}_{1} + (s_{n^{2}-n+1}s^{\prime}_{1} +
s_{n^{2}-n+2}s^{\prime}_{2} + ... + s_{n^{2}}s^{\prime}_{n})
\nonumber
\\ +
(s_{1}s^{\prime}_{1}+ ...+s_{n^2}s^{\prime}_{n^2}) \}] \nonumber
\\ \equiv \exp[\beta {\cal{T}}_{ji}].
\label{transfer}
\end{eqnarray}
In higher dimensional generalizations, ${\cal{T}}_{ji}$ will
be slightly more nested  with the span
$n^{2}$ replaced by $n^{d-1}$.
In Eqn.(\ref{transfer}), the matrix indices $i$ and $j$  
are written in binary 
numerals in terms of the 
$n^{2}$ spins
\begin{eqnarray}
j = \sum_{\alpha=1}^{n^{2}} (s_{\alpha}+1) 2^{\alpha-2}
\end{eqnarray}
where $\alpha$ is the one dimensional coordinate
along the bra spin indices.  
By Eqn.(\ref{transfer}), 
 ${\cal{T}}_{ji}$ are expressed in 
terms of a sum over two digit products where the
digits are those that appear in the binary 
(length $n$ spin) representation of 
the coordinates $j$ and $i$.

As usual, the partition function 
\begin{eqnarray}
Z = Tr[T^{N}] = \sum_{i} \lambda_{i}^{N}
\end{eqnarray}
where $\{ \lambda_{i} \}$ are
the eigenvalues of the transfer
matrix. In the following $\lambda_{\max}$ 
will denote the largest 
eigenvalue.

Note that the transfer matrix
eigenvalues correspond to 
periodic boundary conditions
as strictly required: 
we employed
{\em translational invariance}
to write the partition 
function expansion in 
Fourier space. If the 
boundary conditions are
not periodic all this 
is void.

By the lack of interference 
effects, the connected diagrams
for the free energy are  
correct to ${\cal{O}}(\beta^{n})$ 
for interactions with $m=n^{2}$
(Eqn.\ref{3-d}).
This along with
\begin{eqnarray}
\beta f =
\sum_{p} f_{p} \beta^{p} \equiv - \frac{\ln Z}{N} = - \ln \lambda_{\max}
\end{eqnarray}
(where the last equality holds in the limit of large 
system size $N$)
and a canonical power expansion
having an infinite radius of convergence
\begin{eqnarray}
\lambda_{\max} = \exp[- \sum_{p} f_{p} \beta^{p}]
= \sum_{m=0}^{\infty} \Big( \frac{1}{m!} (- \sum_{p} f_{p}
\beta^{p})^{m} \Big) \nonumber
\\ = \sum_{p} b_{p} \beta^{p}
\label{eig}
\end{eqnarray}
imply that $\lambda_{\max}$ is correct to ${\cal{O}}(\beta^{n})$.

If we
already know the lower
order coefficients 
corresponding to ${\cal{O}}(\beta^{p})$
with $p=0,1,2,3,...,(n-1)$
when we consider the 
$Range= n$ problem. 
then we may set out to compute is
the coefficient of $\beta^{n}$. 
This will amount to extracting
only a single unknown (i.e. the term corresponding
to the coefficient of $\beta^{n}$) from 
a transfer matrix eigenvalue equation,
This can be done to the next
order etc. recursively.
A simple yet
very long single 
linear relation
gives the largest transfer 
matrix eigenvalue to each 
higher order in the 
inverse temperature
$\beta$.

Explicitly, we
can write longhand
\begin{eqnarray}
\det (T- \lambda) = \epsilon_{i_{1}...i_{2^{n^{2}}}}
\prod_{j=1}^{2^{n^{2}}}
[T_{ji_{j}}(\beta)- \lambda(\beta) \delta_{j i_{j}}]= 0
\label{det}
\end{eqnarray}

All one has to do is
to yank out the ${\cal{O}}(\beta^{n})$
term from the sum over products of 
$2^{n^{2}}$ terms. If we pull out a power $p_{j i_{j}}$ 
from each of the elements $[T_{ji_{j}}(\beta)- 
\lambda(\beta) \delta_{j i_{j}}]$ then those will 
need to satisfy 
\begin{eqnarray}
\sum_{j=1}^{2^{n^{2}}} p_{ji_{j}} = n.
\label{powercounting}
\end{eqnarray}


In each element of the transfer
matrix, $\exp[\beta {\cal{T}}_{ji}]$, 
the coefficient of $\beta^{p}$ is
trivially ${\cal{T}}_{ji}^{p}/p!$. 
For each permutation (or ``path'' of coordinates 
$(i,j)$ within the matrix) to be summed 
for the evaluation of the 
determinant

\begin{eqnarray}
[\sum_{path} {\cal{T}}_{ji_{j}}]^{n} = 
\sum_{p_{1},p_{2}, ...., p_{2^{n^{2}}}} \frac{n!}{p_{1}! p_{2}!
... p_{2^{n^{2}}}!} \prod_{\sum p_{ji_{j}} = n}{\cal{T}}_{ji_{j}}^{p_{j}}
\end{eqnarray}
is exactly the product of the coefficients of $\beta^{p_{ij}}$
such that those lead a net power $\beta^{n}$ (i.e. satisfying 
Eqn.(\ref{powercounting}))
from that particular path.

It follows that net contributions 
stemming from $T_{ji_{j}}(\beta)$  in 
Eqn.(\ref{det}) 
are 
\begin{eqnarray}
F^{(n)} \equiv  \frac{1}{n!} \sum_{all ~ paths} \epsilon_{path} [\sum_{(ji_{j}) ~ in ~a ~
given~ path} {\cal{T}}_{ji_{j}}]^{n}
\end{eqnarray}
where $\epsilon_{path}$ simply denotes 
the sign of the given permutation $\{ j \}_{j=1}^{2^{n^{2}}} \rightarrow \{ i_{j}\}$.

To take into account the 
products including 
$\lambda(\beta)$ let us
define  
\begin{eqnarray}
D^{(n;p)}_{2m} = \frac{1}{(n-p)!} 
\sum_{all~ 2m~ paths}  \nonumber
\\  \epsilon_{path} 
[\sum_{(ji_{j}) ~ in ~a ~given~ path} ^{\prime}
{\cal{T}}_{ji_{j}}]^{n-p}
\end{eqnarray}

where the summation is over paths 
going through all possible $(2m)$ 
given points on the diagonal and $\sum^{\prime}$
denotes a summation over ${\cal{T}}_{ji_{j}}$ in the
given paths sans the contributions from
the $(2m)$ diagonal points. 
The determinant paths in $\sum_{all~ 2m~ paths}$ 
can pass through more then $(2m)$ diagonal points-- it is
just that we need to sum over all
those that in their pass also traverse
all possible sets of $(2m)$ given points 
on the diagonal and where those points 
are excluded from the second
${\cal{T}}_{ji_{j}}$
summation.

The $\beta^{n}$ component of Eqn.(\ref{det})
is

\begin{eqnarray}
F^{(n)} + \sum_{p=0}^{n} \sum_{m=1}^{2^{n^{2}-1}}   D_{m}^{(n;p)}
\sum_{\sum_{i=1}^{2m} p_{i} =p} b_{p_{1}} ... b_{p_{2m}} = 0 
\label{rec}
\end{eqnarray}


where $\{b_{p}\}_{p=0}^{n}$ appear in 
the expansion of the largest 
eigenvalue (Eqn.(\ref{eig})).

If the coefficients $ \{b_{p}\}_{p=0}^{n-1}$
then Eqn.(\ref{rec}) where $b_{n}$ appears 
there only linearly gives $b_{n}$ and one may then
proceed to look 
at the eigenvalue equation
for the next value of $n$.

In the absence of
an external field,
for $n=1$ the 
largest eigenvalue for the 
$d$ dimensional problem is 
$\lambda_{+} = 2 \cosh d \beta$
and other eigenvalue
$\lambda_{-} = 2 \sinh d \beta$.
For the three dimensional 
case setting $m=n^{2}=n=1$ 
leads to the one dimensional momentum
space kernel $\hat{v}(k_{1}) = - 6 \cos k_{1}$
which in real space corresponds to 
\begin{eqnarray}
H = - 3 \sum_{i} S_{i} S_{i+1}.
\end{eqnarray}
Of the two eigenvalues only $\lambda_{+}$ has a nonzero $b_{0}$. 
Knowing the values $b_{0}=2$ and $b_{1}=0$ of the largest 
eigenvalue we may
proceed to find $b_{2}$ from 
the Eqn.(\ref{rec}) when $n=2$
etc. This, of course, may be extended
to systems with magnetic
fields.

Setting $b_{0}=2$ in 
Eqn.(\ref{rec}) we find 
\begin{eqnarray}
- b_{n} \sum_{m=1}^{2^{n^{2}-1}}  m 4^{m} D_{m}^{(n;n)} 
  = \nonumber
\\ 
F^{(n)}  + 
\sum_{p=0}^{n-1} 
 \sum_{~p= p_{1}+p_{2}} b_{p_{1}} b_{p_{2}}
D_{1}^{(n;p)} \nonumber
\\ + 
\sum_{p=0}^{n} \sum_{m=2}^{2^{n^{2}-1}}   D_{m}^{(n;p)}
\sum_{\sum_{i=1}^{2m} p_{i} =p~~ ~with~ p_{i} \neq p }
b_{p_{1}} ... b_{p_{n}}. 
\end{eqnarray}
This is the rather explicit recursive relation
giving $b_{n}$ once $\{b_{p}\}_{p=0}^{n-1}$
are known. To find out the coefficient of 
$b_{n}$ let us look at

\begin{eqnarray}
\overline{D}_{s}^{(n;n)} = \sum_{exactly~2s~diagonal~paths} \epsilon_{path},
\end{eqnarray}

where the summation admits only those
permutations that lead to
exactly $2s$ diagonal elements. For the $2^{n^{2}}$ diagonal path 
in the matrix ${\cal{T}}$: $\epsilon_{path}=1$.
For all paths threading $2^{n^{2}-2}$ diagonal
sites $\epsilon_{path}=-1$ and the number
of such paths is ${2^{n^{2}} \choose 2}$.
For the general case the last quantity amounts
to 
\begin{eqnarray}
\overline{D}_{s}^{(n;n)} = (-1)^{s} {2^{n^{2}} \choose 2s}.
\end{eqnarray}
Evaluating, we find
\begin{eqnarray}
D_{m}^{(n;n)} = \sum_{s=m}^{2^{n^{2}-1}} \overline{D}_{s}^{(n;n)}
= 
\frac{(2^{n^{2}})!}{(4m)!(2^{n^{2}}-4m)!}\nonumber
\\ \times 
F_{pq}[\{1,\frac{m}{2}-2^{n^{2}-1},\frac{1}{2}+\frac{m}{2}-2^{n^{2}-1}\},
\nonumber
\\ \{\frac{1}{2}
+ \frac{m}{2},1+2m\},1]
\end{eqnarray}
where $F_{pq}$ is a generalized hypergeometric function.

Insofar as simple geometric visualization 
is concerned, it is amusing to
note, as shown in Fig.(\ref{string}) for
the two dimensional case, that employing 
the usual high temperature
expansion in powers of $\tanh \beta$ one
would reach the same conclusion
regarding the correctness (up to order $n$) 
of the partition function evaluated for 
a width $n$ slab vis a vis the 
partition function of the two 
dimensional system. We 
may look at the a finite thickness $(n$) slab 
of the two dimensional lattice along which we
apply periodic boundary conditions.
Let us now draw a string going 
along one row of length $n$,
after which it would jump to the next row, scan
it for $n$ sites, jump to the next one, and so
on. On the one dimensional laced string,
the system is translationally invariant and
the interactions are of ranges $R=1,n$.
Or, explicitly, by counting the number of 
closed loops in real space
(employing the standard, slightly
different, diagrammatic expansion
in powers of $\tanh \beta$),
we see that the terms
in this expansion are also identical
up to order $\tanh^{n} \beta$ \cite{Triangular}.

A lacing of a two dimensional slab by a one 
dimensional string, as shown in 
the figure, is one of the backbones  
of Density Matrix Renormalization
Group Theory when applied
to two dimensional problems.

\begin{figure}
\begin{center}
 \epsfig{figure=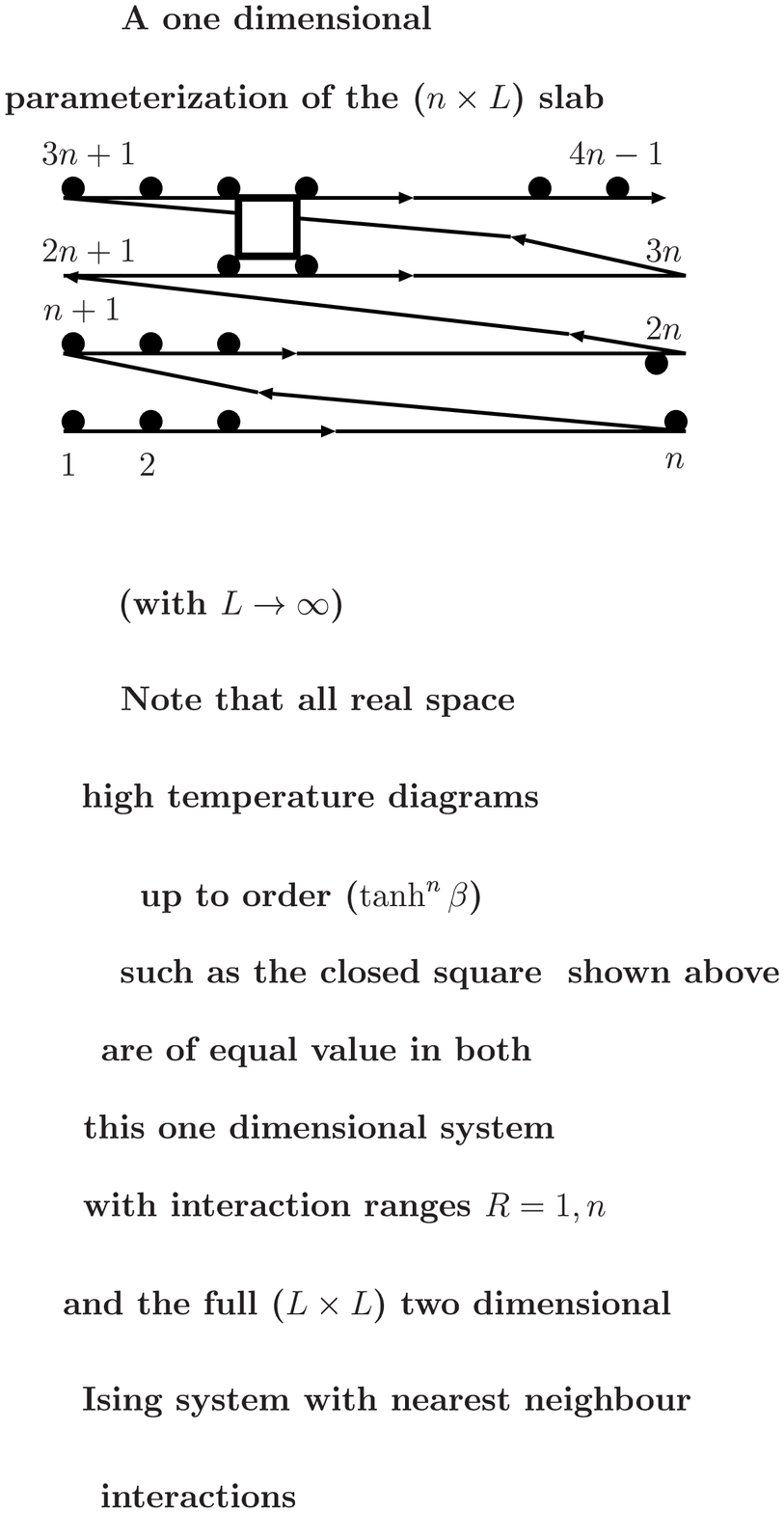,width=0.85\linewidth}
\end{center}
\caption[]{Lacing of a ``two-dimensional'' slab. 
The width of the slab ``$n$'' serves 
as an inverse temperature axis in the following
sense- the wider it is, the higher order in 
$1/T$ that we may advance towards 
the full two dimensional model.
We may get an analogous ``three-dimensional''
slab with ``two imaginary time axis''
if we set $m=n^{2}$. The path will
then thread $n^{2}$ sites before 
continuing upward.}
\label{string}
\end{figure}

\subsection{Incommensurate dimensional reduction} 
\label{incomm}

\bigskip

{\bf Theorem:}
The one dimensional real space kernel
\begin{eqnarray}
\hat{V}(x_{1}) = - 2  \sum_{l=1}^{d} 
[(c_{l}+x_{1}) \sin \pi(c_{l}-x_{1}) \nonumber
\\
+ (c_{l}-x_{1})] \sin
\pi(x_{1}+c_{l})]\nonumber
\\ \times [(c_{l}-x_{1})(c_{l}+x_{1})]^{-1}, 
\label{redu}
\end{eqnarray}
(i.e. a scenario in which each
spin is effectively composed of $d$ shifted ``Coulombic sources''
(more precisely, each spin is 
composed of $d$ pairs of ``charges'' 
generating sinc potentials))
will give rise to the exact 
$d$ dimensional nearest neighbor 
partition function and 
free energies when those
quantities are averaged 
over $\{c_{l}\}_{l=1}^{d}$.

Once again the proof is not too
involved. The basic idea is that
in momentum space 
this will give rise to
\begin{equation}
\hat{v}(k_{1})= - 2 \sum_{l=1}^{d} \cos c_{l} k_{1}
\end{equation}
which when averaged over 
incommensurate $\{c_{l}\}$
will annihilate all phase
coherent terms and
reproduce the expansion 
with the $d$ dimensional 
kernel in Eqn.(\ref{d-dim}).

Many different measures for
$d \mu (c_{l})$ can be chosen.
Perhaps the simplest 
one is 
\begin{equation}
\frac{1}{2 \Lambda} \int_{-\Lambda}^{\Lambda} d c_{l} 
\end{equation}
for each of the d coefficients $\{ c_{l} \}$. 

 Performing the 
$\{ c_{l}\}_{l=1}^{d}$ integrations after
the loop integrals over 
$\{ k_{a} \}$ we will find that
the most general integral is 
of the form 
\begin{eqnarray}
\frac{1}{\Lambda^{d}}
\prod_{l=1}^{d} \int_{-\Lambda}^{\Lambda}  
dc_{l} \nonumber
\\ ~ ~ \Big[ \prod_{b} \int_{-\pi}^{\pi} \frac{dk_{b}}{2 \pi}
\prod_{a} \sum_{l=1}^{d}   
\exp[i \sum_{b} M_{ab} k_{b} c_{l}]~ \Big]
\end{eqnarray}
where the index $b$ runs over the various 
independent loop momenta
($k$ is still merely a 
scalar for this one dimensional
problem). Unless, for a given $k_{b}$ integration,
the argument of the exponent is identically
zero (i.e. corresponding to
a term that would be generated 
in the d-dimensional nearest neighbor
problem) an ``interference term'' results.
However, such a term is down by ${\cal{O}}(\Lambda^{-1})$ 
by comparison to the ``good'' noninterference 
terms that occur in the d-dimensional
problem. For each given assignment
of $\{c_{l}\}$ along the propagator lines
we may perform some coefficient averages  
over a few of the $\{ c_{l}\}$ first and then integrate over
the loop momenta and average over the 
remaining coefficients. For ``bad'' interference terms,
i.e. in those cases in which the argument of the exponential is not
identically zero prior to the integration, the 
canonical integral
\begin{equation}
\frac{1}{2 \Lambda} \int_{-\Lambda}^{\Lambda} dc \exp[ic k] \rightarrow
\frac{\pi \delta(k)}{\Lambda}
\end{equation}
will result in the $\Lambda \rightarrow \infty$ limit.
When such a delta function
is integrated over the momenta in the loop integrals
the resulting term is ${\cal{O}}(\Lambda^{-1})$
down by comparison to a noninterference
term for which the integral 
would read
\begin{equation}
\frac{1}{2 \Lambda} \int_{-\Lambda}^{\Lambda} dc (1).  
\end{equation}

In the limit $\Lambda \rightarrow \infty$
such ``bad'' interference 
terms will evaporate
both in the connected
diagrams (for the 
free energy calculation)
as well as for the 
disconnected diagrams
(included in the 
evaluation of
the partition
function). 

Q.E.D.

\bigskip

{\bf Corollary:}

For a single spin 1/2 particle
with an action 
\begin{eqnarray}
S = \int_{0}^{\beta} d {\tau} 
\int_{0}^{\beta} d \tau^{\prime} ~ \overline{\psi}(\tau)
\hat{V}(\tau-\tau^{\prime}) \psi(\tau^{\prime}) \nonumber
\\ = \sum_{\omega_{n}}  \overline{\psi}(-\omega_{n}) \hat{v}(\omega_{n}) \psi(\omega_{n})
\end{eqnarray}
where $\psi(\tau)$ is the two component
spinor, the averaged partition function
is identically the same as that of 
the $d$ dimensional nearest
neighbor ferromagnet.

The proof of this
statement trivially
follows from breaking up
the $[0,\beta]$ segment
on the imaginary time axis
into $L$ pieces and allowing 
$L \rightarrow \infty$. 

We have mapped the entire three
dimensional Ising model onto
a single spin 1/2 quantum particle!

\bigskip

\bigskip

Employing the Hubbard Stratonovich for the 
quantum case we may make 
some formal one dimensional 
reductions.

\bigskip

\bigskip

{\bf Bosonization In High Dimensions By An Exact Reduction To One Dimension}

Fermionic electron (or other) fields may be
formally bosonized on each individual chain.
The one dimensional band,  
\begin{eqnarray}
\hat{v}_{1} = -2 \sum_{l=1}^{d} \cos c_{l}
k_{1},
\label{one-d-}
\end{eqnarray}
is now a simple sum of $d$ cosines (au lieu of the 
standard single tight binding cosine). For each given 
set $\{c_{l}\}_{l=1}^{d}$, the band dispersion of Eqn.(\ref{one-d-}) 
may be easily linearized about 
its two respective Fermi points. Consequently 
the standard bosonization methodology 
may be applied. Averaging over $\{c_{l}\}_{l=1}^{d}$
with various weights (corresponding to the different observables) 
leads to the corresponding d-dimensional quantities. 

\bigskip

{\bf Jordan Wigner Transformation}

On mapping the three dimensional 
Heisenberg model to 
a spin chain (more precisely average over
spin chains)  with sinc like 
interactions, the Jordan-Wigner
transformation may be 
applied. On the chain we may set
\begin{eqnarray}
\sigma_{i}^{z} = 1 - 2 a_{i}^{\dagger} a_{i} \nonumber
\\ \sigma^{+}_{i} = \prod_{j<i} (1- 2 a_{j}^{\dagger} a_{j}) a_{i}
\nonumber
\\ \sigma^{-}_{i} = \prod_{j<i} (1- 2 a_{j}^{\dagger} a_{j})
a_{i}^{\dagger}
\end{eqnarray}
where $\sigma_{i}^{\pm}=(\sigma_{i}^{x}\pm i \sigma_{i}^{y})/2$.
and the $\{a_{i}\}$ operators satisfy Fermi statistics
\begin{eqnarray}
\{ a_{i},a_{j}^{\dagger} \} = \delta_{ij}, \{a_{i},a_{j}\} =
\{a_{i}^{\dagger},a_{j}^{\dagger}\} =0.
\end{eqnarray}
Though now the Jordan Wigner 
can be effected to any translationally 
invariant high dimensional problem, the 
resulting fermion problem 
is in general very complicated.

\section{Permutational symmetry}
\label{permutational}

The spherical model (or  $O(n \rightarrow \infty)$) 
partition function
\begin{equation}
  Z = const \left(\prod_{\vec{k}}
    \left[\frac{1}{\sqrt{\beta[\hat{v}(\vec{k})+\mu]}}\right]\right),
\end{equation}
where the chemical
potential
$\mu$ satisfies
\begin{eqnarray}
\beta = \int \frac{d^{d}k}{(2 \pi)^{d}} \frac{1}{\hat{v}(\vec{k}) + \mu},
\end{eqnarray}
is invariant under permutations of $\{\hat{v} ( \vec{k})\} \rightarrow
\{\hat{v}(P\vec{k})\}$. In the above, the permutations
\begin{eqnarray}
\{\vec{k}_{i} \}_{i=1}^{N} \rightarrow \{ P \vec{k} \}
\end{eqnarray}
correspond to all possible shufflings of the 
$N$ wavevectors $\vec{k}_{i}$.

This simple invariance allows all d-dimensional 
translationally invariant systems to 
be mapped onto a 1-dimensional one. Let us design an effective
one dimensional kernel $V_{eff}(k)$ by
\begin{equation}
  \int \delta[\hat{v}(\vec{k})-v] d^{d}k =
  |\frac{dV_{eff}}{dk}|_{V_{eff}(k)=\hat{v}}^{-1}.
\end{equation}
The last relation secures that
the density of states and consequently
the partition function is preserved.  For 
the two-dimensional nearest-neighbor ferromagnet:
\begin{eqnarray}
  |\frac{dV_{eff}}{dk}|^{-1} =\rho(V_{eff}) \nonumber
\\ = c_{1}
  \int_{0}^{1}\frac{dx}{\sqrt{1-x^{2}}\sqrt{1-(V_{eff}+x-2)^{2}}},
\end{eqnarray}
and consequently 
\begin{eqnarray}
  k(V_{eff})= c_{1}\int_{0}^{V_{eff}}
  F(\sin^{-1}\sqrt{\frac{2}{(3-u)u}},\frac{\sqrt{4u- u^{2}}}{2}) du,
\label{elliptic}
\end{eqnarray}
where $F(t,s)$ is an incomplete elliptic integral of the first kind.
Eqn.(\ref{elliptic}) may be inverted
and Fourier transformed to find the effective one dimensional 
real space kernel $\hat{V}_{eff}(x)$.
We have just mapped the two dimensional nearest 
neighbor ferromagnet onto a one dimensional 
system.
In a similar fashion, within the spherical 
(or equivalently the $O(n \rightarrow \infty)$) limit
 all high dimensional problems may be mapped
onto a translationally invariant one dimensional 
problem. It follows that 
the large $n$ critical exponent of the
$d$ dimensional nearest neighbor ferromagnet
are the same as those of translationally invariant 
one dimensional system with longer range
interactions. We have just shown 
that a two dimensional $O(n \gg 1)$ 
system may has the same thermodynamics
as a one dimensional system.
By permutational symmetry,
such a maping may be performed
for all systems irrespective
of the dimensionality of the 
lattice or of the nature of the 
interaction (so long
as it translationally
invariant). We have just demonstrated
that the notion of universality 
may apply only to the canonical 
interactions.

The lowest order term
breaking permutational symmetry in our high temperature
expansion is $\eta^4(\vec{x})\eta^{4}(\vec{y})$.  
Thus permutational symmetry is broken 
to ${\cal{O}}(\beta^{4})$ for finite $n$.
For a constraining term (e.g. $\sum_{\vec{x}} \ln[\cosh[\eta(\vec{x})]]$ for
$O(1)$ spins) symmetric in $\{\eta(\vec{k})\}$ to a given order, one
may re-arrange the non-constraining term $\sum_{\vec{k}}
\hat{v}^{-1}(\vec{k}) |\eta(\vec{k})|^{2} = 
\sum_{\vec{k}} \hat{v}^{-1}(P \vec{k})
|\eta(P \vec{k})|^{2}$) and relabel the dummy integration variables
$H[\{\eta(\vec{k})\}] \rightarrow H[\{\eta(P^{-1}\vec{k})\}]$ to effect
the constraining term augmented to a shuffled spectra $\hat{v}
(P \vec{k})$.

{\em Acknowledgments.}

This research was supported by the Foundation
of Fundamental Research on Matter (FOM), which is sponsored by the
Netherlands Organization of Pure research (NWO).

{\bf References}\\


\begin{thebibliography}{99}
\bibitem{N}
As a technical aside, note that 
on the lattice (unlike the standard
continuum theory) each vertex 
conserves momentum only
up to a reciprocal lattice 
vector $\vec{K} = 2 \pi (n_{1},n_{2}, ..., n_{d})$
with $n_{i}$ integers.
Formally, this is as $\vec{x}$ only take on discrete
values on a lattice,

\begin{equation}
\sum_{\vec{x}} \exp[i \vec{k} \cdot \vec{x}] = N \sum_{\vec{K}}
\delta_{\vec{k},\vec{K}}
\end{equation}
where $\vec{K}$ are reciprocal lattice 
vectors. 
 
Thus at each vertex of order $2m$ we have 
\begin{equation}
\frac{1}{(2 m)!} \sum_{\vec{K}} \delta(\vec{q}_{in}-\vec{K},0)
\end{equation}
where $\vec{q}_{in}$ denotes the net momenta 
flowing into the vertex. As for all lattice 
models the propagators are periodic 
$\hat{v}(\vec{q}) = \hat{v}(\vec{q}+\vec{K})$
for each of the $N$  reciprocal lattice vectors $\vec{K}$.
As such the momentum conservation constraint 
may be replaced
by 
\begin{equation}
\frac{1}{(2 m)!} \sum_{\vec{K}} \delta(\vec{q}_{in}-\vec{K},0) 
\rightarrow N \frac{1}{(2 m)!}  \delta(\vec{q}_{in},0). 
\end{equation} 
In the aftermath, each vertex carries a factor
of the volume ($N$) multiplying the discrete
Kronecker delta function. After the momentum
conservation constraints are taken care of
at each vertex, we will be left
with a summation over all remaining independent loop
momenta $\{ \vec{k}_{b} \}$. Each one of these 
summations is to be replaced by the standard
\begin{equation}
\sum_{\vec{k}} \rightarrow N \int_{-\pi}^{\pi} ...  
\int_{-\pi}^{\pi} \frac{d^{d}k}{(2 \pi)^{d}}. 
\end{equation}
Thus, at the end we will be left with 
a factor of
\begin{equation}
N^{\mbox{loops + vertices - propagator~ lines}}
\end{equation}
multiplying an integral over the independent loop
momenta $\{ \vec{k}_{b} \}$ in Eqn.(\ref{loop}).
Compounding all together, as in each closed diagram
the number of independent loops + number of
delta function momentum conserving vertices  =
1 + number of propagator lines, a net factor of $N$
should be attributed to each independent closed
bubble just as it does in the 
conventional continuum theories.
\bibitem{Gibbs} 
As noted earlier, for any
lattice model 
\begin{equation}
\hat{v}(\vec{k}) = \hat{v}(\vec{k}+\vec{K}),
\end{equation}
as $\exp[i \vec{K} \cdot \vec{x}]=1$ in 
\begin{eqnarray}
\hat{v}(\vec{k})= \sum_{\vec{x}} \hat{V}(\vec{x}) 
\exp[i \vec{k} \cdot \vec{x}].
\end{eqnarray}
The sum over $\vec{x}$ spans all
discrete lattice sites. 


The way to analytically continue
$v(\vec{k})$ when $k_{1}$ becomes complex is 
to make the $k_{2}$ dependence also 
invariant under shifts 
by $(2 \pi)$. Within each Brillouin
Zone the $k_{2}$ dependence stems
from the analytic continuation
of the functional 
dependence on 
the real part.
The only complication
is the Gibbs' phenomenon
in $k_{3}$ which occurs 
when taking the Fourier 
transform of the lattice 
$\hat{V}(\vec{x})$.
This is slightly 
avoided for the real
part of $V(\vec{x})$ as follows:
Under parity  and
all individual 
reflections:
$\hat{V}(x_{1},x_{2},x_{3}) = \hat{V}(x_{1}, x_{2},-x_{3})$
etc. This implies that in the 
Fourier transform $\hat{v}(\vec{k})$: ~
$\hat{v}(k_{1},k_{2},k_{3}) = \hat{v}(k_{1},k_{2},-k_{3})$.
Thus, taking only the even part in $k_{3}$
of $\cos(k_{2}+ i \lambda k_{3})$ we will only
be left with the terms $\cos k_{2} \cosh k_{3}$.
Now, $\cosh k_{3}$ is continuous and periodic 
when the Brillouin Zone is 
chosen as $ -\pi < k_{3} \le \pi$. 
No Gibbs phenomenon occurs. The left and 
right derivatives are still discontinuous 
at the zone edges.
\bibitem{Triangular}
It might be easier to compute the terms
in the $\beta$ expansion for a
three dimensional Ising nearest neighbor
ferromagnet by going to 
a triangular lattice and
deleting all
bad ``constructive interference''
loop integrals. 
A line threading the triangular lattice will
connect a spin with spins of distance
$Range=1,$ $n,$ and $n+1$ 
along the line. 
``Bad interference terms'' can result
from loop integrals
of the form $(\exp[ik]  \exp[ink] \exp[-i(n+1)k])^{p}$.
This is the only way in which loop integrals
can conspire to give nonzero results
when $n \rightarrow \infty$.
Thus, trivially, the 
high temperature expansion
for $d=3$ cubic ferromagnet
is equivalent
to that of the triangular
ferromagnet up to 
order $\beta^{3-1}$.
Similarly, if we set 
$s=n-1$ we obtain a
2-Dimensional lattice
model in which each 
spin interacts with
its four nearest neighbors
and four diagonal neighbors.
\bibitem{ }
Employing $\beta f= - (\ln Z)/N = - 2 \cosh \beta$ 
for the free energy density in the canonical
one dimensional case trivially leads to the 
high temperature coefficients:
\begin{eqnarray}
f_{p} = - \frac{2}{(2m)!}.
\end{eqnarray}  
This is the average of 
the coefficients in 
high temperature expansion
in the kernel is the long ranged
one of Eqn.(\ref{redu}) with
$d=1$. Similarly, if 
we set $d=2$ in  Eqn.(\ref{redu})
and average we will find the 
high temperature coefficients
of the two dimensional
nearest neighbor Ising
model.

Just as trivially, if we examine long ranged interactions
in two dimensional spawned by 
\begin{eqnarray}
\hat{V}(x_{1},x_{2}) = - 2 \Big( \delta_{x_{2},0}
[(c_{1}+x_{1}) \sin \pi(c_{1}-x_{1}) \nonumber
\\
+ (c_{1}-x_{1})] \sin
\pi(x_{1}+c_{1})]\nonumber
\\ \times [(c_{1}-x_{1})(c_{1}+x_{1})]^{-1} \nonumber
\\ 
+ \delta_{x_{1},0}
[(c_{2}+x_{2}) \sin \pi(c_{2}-x_{2}) \nonumber
\\
+ (c_{2}-x_{2})] \sin
\pi(x_{2}+c_{2})] \nonumber
\\ \times [(c_{2}-x_{2})(c_{2}+x_{2})]^{-1} \Big)
\label{red-pro}
\end{eqnarray}
 
- the Fourier transform of 
\begin{eqnarray}
\hat{v}(\vec{k}) = -2 (\cos c_{1} k_{1} + \cos c_{2} k_{2}),
\end{eqnarray}

then we will find that 
the average over $c_{1}$ and $c_{2}$ 
of the coefficients in 
the resulting high temperature series
for this two dimensional 
long ranged kernel will
have the coefficients 
of the high temperature
expansion of the two dimensional
nearest neighbor Ising 
model.
\bibitem{antiperiodic}

Note the unmentionably obvious: 
for antiperiodic boundary
conditions (as in a 
fermionic problem along
the imaginary time 
$\tau$ axis) a Fourier
space expansion is 
again possible however
only the Matsubara frequencies
$\beta \omega_{n} = 2 \pi (n+1/2)$ 
with integer $n$ will
appear.
\bibitem{ }

The Coulomb charges take on the role
of the nearest neighbor interactions
on the original d dimensional lattice.
If one would naively thread a line through the d-dimensional
lattice (such that net number of points and the spin spin
interactions are preserved) then one would find integer
long range interactions on that line. An integer
range interaction becomes a shifted (by the same
range) sinusoidal Coulomb like
interaction in the continuum. The last statement
follows from Fourier transforms.
\end{thebibliography}
\end{document}